# Postprint

This is the accepted version of a paper published in Scientometrics. This paper has been peer-reviewed but does not include the final publisher proof-corrections or journal pagination.

Access to the published version may require subscription.

N.B. When citing this work, cite the original published paper.

# Research assessment under debate: disentangling the interest around the DORA Declaration on Twitter

Enrique Orduña-Malea[1]  and Núria Bautista-Puig[2]

[1] Department of Audiovisual Communication, Documentation and History of Art, Universitat Politècnica de València, Valencia (Spain)
✉ enorma@upv.es

[2] Department of Library and Information Sciences, Universidad Complutense de Madrid, Madrid (Spain)
✉nuriabau@ucm.es
*Corresponding author

## Abstract

Much debate has been around the misapplication of metrics in research assessment. As a result of this concern, the Declaration on Research Assessment (DORA) was launched, an initiative that caused opposing viewpoints. However, the discussion topics about DORA have not been formally identified, especially in participatory environments outside the scholarly communication process, such as social networks. This paper contributes to that end by analyzing 20,717 DORA-related tweets published from 2015 to 2022. The results show an increasing volume of tweets, mainly promotional and informative, but with limited participation of users, either commenting or engaging with the tweets, generating a scarcely polarized conversation driven primarily by a few DORA promoters. While a varied list of discussion topics is found (especially "Open science and research assessment," "Academics career assessment & innovation," and "Journal Impact Factor"), the DORA debate appears as part of broader conversations (research evaluation, open science). Further studies are needed to check whether these results are restricted to Twitter or reveal more general patterns. The findings might interest the different evaluators and evaluated agents regarding their interests and concerns around the reforms in the research evaluation.

# 1 Introduction

Supporting the quality of research and strengthening societal trust in the research and innovation systems (including their outputs) should be the backbone for researchers, research organizations, and research funders to achieve excellence (*European Commission*, 2021). Consequently, how science and researchers are assessed is crucial. However, traditional assessment processes rely predominantly on journal-based quantitative metrics, promoting the so-called 'publish or perish' culture (Van Dalen & Henkens, 2012), which might have a pernicious effect on academics' lives, especially on funding, tenure, promotion, and recruitment (Moher et al., 2018). For this reason, these academic evaluation regimes have been criticized for being widely oriented to metrics, harming academic environments and knowledge production, and promoting deceptive actions (Kulczycki, 2023).

The debate around improving research assessment systems goes back decades, with a broad agreement on the need for research assessment reform to further support research quality and the researchers' value (Delgado-López-Cózar, Ràfols & Abadal, 2021). Among the issues that have led to the debate about research assessment, there is a variety of interconnected reasons, such as the misuse of performance indicators (Moed, 2007; Wildgaard, 2015), especially the Journal Impact Factor (Lariviere & Sugimoto, 2019), the unintended effect consequences that the use of evaluation systems has had, such as gift authorship (Abramo, D'Angelo & Di Costa, 2019), the strategic behavior and goal displacement (Abramo, D'Angelo & Grilli, 2021; Akbaritabar, Bravo & Squazzoni, 2021) or other deceptive practices oriented to game the system (Kulczycki, 2023).

In addition, epistemic injustices such as the language of publications (Rowlands and Wright, 2020), global-local competition (Vidovich, 2008), the use of skewed bibliographic databases with adverse effects in some fields (Mongeon & Paul-Hus, 2015; Martín-Martín et al., 2021) or the lack of recognition of plural research outputs in support to diverse and inclusive research culture, such as interdisciplinarity, gender, or diversity (Northcott & Linacre, 2010), have been highlighted by the literature.

Other debates have been related to finding an optimal solution, such as the combination of quantitative and qualitative approaches (Van Raan, 2005) or proposing new alternative indicators to overcome these limitations (Priem & Hemminger, 2010), with their advantages and disadvantages (Thelwall, 2020).

Specific recommendations to improve research evaluation have increasingly come from various sources and expert groups. The Declaration on Research Assessment- DORA (2012), the Leiden Manifesto for Research Metrics (2015), and the Coalition for Advancing Research Assessment (CoARA) agreement (2022), among others, have become at the center of attention in these debates. Most of these initiatives provide recommendations on assessment processes and indicators aimed at different stakeholders (e.g., DORA, Leiden Manifesto). In contrast, others embed the assessment process within the open science spectrum (e.g., The EU's Open Science policy).

Some of these recommendations are available online on websites (e.g., DORA). In contrast, others have been published as technical reports (e.g., Next Generation Metrics [*European Commission* et al., 2017], the Metric Tide [Wilsdon et al., 2015]) or scientific articles (e.g., Leiden Manifesto [Hicks et al., 2015], Hong Kong Principles for Assessing Researchers [Moher et al., 2020]).

Arguably, the best-known of these initiatives is DORA. Most studies and editorial letters just mention the Declaration (see Moher et al., 2018), whereas others cite its benefits to one field (see O'Connor, 2022), support the Declaration (see Welk et al., 2014; Parish et al., 2018) or analyze the progress and challenges after "x" years since its announcement (see Schmid, 2017).

The greater or lesser acceptance and implementation of DORA affects the functioning of the scientific community since its adoption implies changes in regional, national, and supranational scientific policies, the operation of evaluation agencies, and the promotion of research personnel in universities and research centers. Therefore, describing and characterizing the discussion topics around DORA is relevant to understanding the concerns, problems, and challenges faced by different actors in the scientific ecosystem.

Although the debate for or against DORA (and other manifestos) has been widely discussed in the scientific literature (see Section 2), the study of the discussion generated around DORA outside of scholarly communication (especially on social networks) is of particular relevance to carry out a more granular analysis of DORA, identifying thus specific discussion topics about this declaration that might remain invisible in scholarly publications. This way, it is possible to find out whether this debate attains attention around particular issues (e.g., use of the Journal Impact Factor) but not around others (e.g., outputs of scientific research) or whether these topics are linked to other movements (e.g., open science). In addition, we can also determine the professional participation of journals, institutions, and practitioners in the debate beyond individual scholars.

The main objective of this study is to identify and characterize the discussion topics around DORA on Twitter. The following research questions are established to facilitate the achievement of these objectives:

**RQ1**. What is the volume of discussion about DORA on Twitter?
**RQ2**. Who is discussing DORA on Twitter?
**RQ3**. Which topics about DORA generate interest on Twitter?
**RQ4.** Do topics about DORA on Twitter evolve?
**RQ5**. What sentiment do the tweets about DORA generate?
**RQ6**. What impact do the tweets about DORA generate?

## 2  Background

The literature on research assessment is broad and thematically diverse. Pérez-Esparrells, Bautista-Puig & Orduña-Malea (2022) highlight the following topics treated under this body of literature: Bibliometric indicators (i.e., compilation of indicators used in evaluation processes, proposals for the improvement of existing indicators, and the design of new indicators); Evaluation systems (i.e., describing the advantages and disadvantages of different research evaluation systems); Bibliographic databases (i.e., description of the databases used as sources of indicators for evaluation processes); Altmetrics (i.e., use of alternative metrics in evaluation processes); Assessment methodologies (addressing the advantages and disadvantages of the use of quantitative and qualitative indicators); Opportunistic behaviour (identification of malpractices among research staff, and its consequences); Cross-cutting issues (inclusion and assessment of other complementary aspects, such as interdisciplinary, gender mainstreaming, diversity, sustainability or local impact); and Recommendations, declarations, or principles, such as DORA.

DORA was created by a group of editors and publishers of scholarly journals during the Annual Meeting of The American Society for Cell Biology (ASCB) in San Francisco, CA, on 16 December 2012, becoming the first Declaration that led to a global and extensive debate on academic research. Signatures up to June 2022 were at 20,375 individuals and 2,854 organizations in 161 countries and has raised community awareness, started meaningful discussions, changed research policies, and generated a large body of scientific literature (Appendix A includes an institutional co-authorship network about DORA for illustrative purposes).

DORA developed a set of recommendations that include the need to eliminate the use of journal-based metrics in funding appointment and promotion, assess research on its own merits (rather than the journal in which the study is published), and capitalize on the opportunities provided by online opportunities (e.g., exploring new indicators of significance and impact). Even though DORA was widely celebrated by a large part of the research community, it addresses a strong attack against misusing and misapplying bibliometrics (Torres-Salinas, Arroyo-Machado & Robinson-Garcia, 2023). This position has been ultimately continued by the launching of the Agreement on Reforming Research Assessment (CoARA) (Sivertsen & Rushforth, 2022). This issue might give rise to the bipolarity of opinions and debate. For that reason, by monitoring the contents and views about DORA, we can help understand the attitudes of the users (individual authors, practitioners, or institutions) toward this declaration.

Twitter is essential in discovering scholarly information (Mohammadi et al., 2018) and is also used to discuss and disseminate scientific outcomes. Beyond the Altmetrics-oriented studies based on the academic use of Twitter by researchers, other studies focus on analyzing the users' opinion on specific science-related topics, such as the Open Access movement (Sotudeh et al., 2022; Sadiq & Yadav, 2022; Sotudeh, 2023), the h-index (Thelwall & Kousha, 2021), the Kardashian Index (Powell, Haslam & Prasad, 2022) or scientific misconduct (Copiello, 2020). However, no studies on analyzing the conversation about DORA have been located, the aspect on which the present research focuses.

## 3 Methods

This article presents a case study based on the analysis of the DORA movement, both directly through DORA's main account (@DORAssessment) and indirectly through the Twitter community mentioning DORA's Twitter account or using DORA-related hashtags.

**Tweets data collection**

The Academic Twitter API was used to collect all tweets published by the DORA official account (hereinafter, the DORA dataset) using the following query: "from:@DORAssessment." This account was created by @ASCBiology in 2015 and accounts for 15 thousand followers when writing this manuscript (June 2023).

In addition, all tweets mentioning DORA's official Twitter account were retrieved through the following query: "@DORAssessment -is:retweet" (from now on referred to as the user-mention dataset). As relevant tweets might not directly mention @DORAssessment, all tweets containing the

#sfDORA, #SFDORA_Declaration, and #DORADeclaration hashtags were also retrieved (from now on referred to as the hashtag-mention dataset). The #SanFranciscoDeclaration hashtag was also used, excluding those tweets related to the "High-Level Policy Dialogue on Women and the Economy," also called the San Francisco Declaration. The $DORA hashtag was banned as it is also used to mention a cryptocurrency, the Digital Operational Resilience Act (DORA), and the name of a television series (DORA the Explorer), thus introducing noise in the query.

The data collection covers the launch of DORA's Twitter account (24 April 2015) to 31 May 2022. Data extraction was carried out by June 2022. For each tweet, the following parameters were collected: author-id, publication date, tweet text, and public metrics (retweets, replies, likes, and quotes). The process yielded a total of 20,807 tweets: 5,977 direct tweets for the DORA dataset, 13,985 tweets for the user-mention dataset, and 845 tweets for the hashtag-mention dataset (the three datasets are disjoint sets of tweets without duplicates). All tweets were later cleaned (e.g., removing stop words, punctuation, URLs, and monograms) by using R v.4.2.0 (*R Core Team*, 2022) along with the following libraries: *tidytext* (Silge & Robinson, 2016), *dplyr* (Wickham et al., 2022), *stringr* (Wickham, 2022) and *stopwords* (Benoit, Muhr, & Watanabe, 2021).

**User data categorization**

The User endpoint of the Academic Twitter API was used to obtain descriptive information for each user (username, followers, and total published tweets). The open dataset of scholars on Twitter (Mongeon, Bowman, & Costas, 2022) was then used to test whether users mentioning DORA or using DORA-related hashtags were scholars, considering the last available version (2022/08/21). This dataset includes 498,672 unique author-tweeter pairs and is the most comprehensive list to identify scholars on Twitter.

For those users with the most published tweets in the datasets, their accounts were accessed manually, and based on the public information indicated, the type of account (personal or institutional) was recorded. In the case of personal accounts, the gender and the role (researcher, practitioner) were registered.

**Tweets data categorization**

A manual and automatic categorization of tweets was performed to determine the discussion topics about DORA.

*Manual categorization*

Tweets and hashtags were categorized. In both cases, the first author conducted a manual inspection to identify the significant topics and design a preliminary categorization. After that, the second author completed a second round, reclassifying when necessary. A third round was carried out after fully agreeing on the results, adjusting the categories, and resolving doubtful cases.

In the case of tweets, a manual inspection of the 20,807 tweets collected separated those tweets related to DORA the organization from DORA the declaration. Given the complexity of this task, the authors avoided frequency calculation and decided to focus on identifying specific topics of discussion (categories). Tweets will be mentioned through their Tweet-ID.

Regarding hashtags, 1,554 were analyzed from 5,280 tweets (26.5% of all tweets included in the DORA and user-mention datasets). Given that the tweets in the hashtag-mention dataset are determined by the presence of a set of DORA-related hashtags, this dataset was excluded from this method as it was biased towards the presence of a few hashtags. In any case, the small size of this dataset (845 tweets) allows its exclusion without significantly affecting the results.

*Automatic classification*
*CorTexT Manager* (Breucker et al., 2016) was used to identify the topics of each tweet. For this purpose, single and multi-terms (n-grams) were extracted from each tweet using a lexical extraction tool based on Natural Language Processing (NLP) techniques. This tool incorporates its own automated method to identify significant terms with high "unithood" (Frantzi, Ananiadou & Mima, 2000) and "termhood" (Kageura & Umino, 1996)[1] to address the time-consuming nature of term extraction within NLP tools. This study computed the specificity for ranking terms using the direct similarity *chi2score* measure (Pearson, 1900), which considers the number of co-occurrences for each pair analyzed. Additionally, different frequency ranges of terms were employed (e.g., 100 nodes for those keywords with a frequency higher than 12 and 300 nodes for those hashtags with a frequency higher than 3). Last, the authors revised and validated the topics identified by the tool.

*CorTexT*'s network mapping tool was used to build the co-occurrence network of topics, which was subsequently processed with Gephi (v. 0.10.1). The spatialization used by this software is the classical Fruchterman-Reingold layout (1991) and the algorithm used for community detection was the Louvain resolution (Blonden et al., 2008).

**Tweet sentiment**

A final dataset of 15,354 English tweets (4,663 from the DORA dataset, 10,379 from the user-mention dataset, and 312 from the hashtag-mention dataset) was included in this analysis (other languages were excluded from the sentiment analysis due to accuracy constraints). Otherwise, 1,592 tweets containing RT at the beginning of the tweet were filtered out to avoid sentiment inflation (several had the same tweet content).

The *CorTexT Sentiment Analysis* tool used a sentiment range from -10 (negative) to +10 (positive) through the Python library *textblob*. The authors tested other sentiment tools, whose results are described in the supplementary material (Appendix B).

## 4  Results

**RQ1. What is the volume of discussion about DORA on Twitter?**
The cumulative monthly frequencies of tweets throughout the study (2015-2022) are displayed in Fig.1. Although DORA's official account was created in 2015, the number of tweets did not start rising until 2018, when the maximum relative frequency was recorded (1,568 tweets by DORA's dataset and 3,508 by the user-mention dataset). Otherwise, the hashtag-mention dataset only represents 4% of all tweets (845). However, it started long before the other datasets (208 tweets in 2013), probably because Twitter users had already commented DORA even though the official account did not exist.

---
[1] https://docs.cortext.net/lexical-extraction/

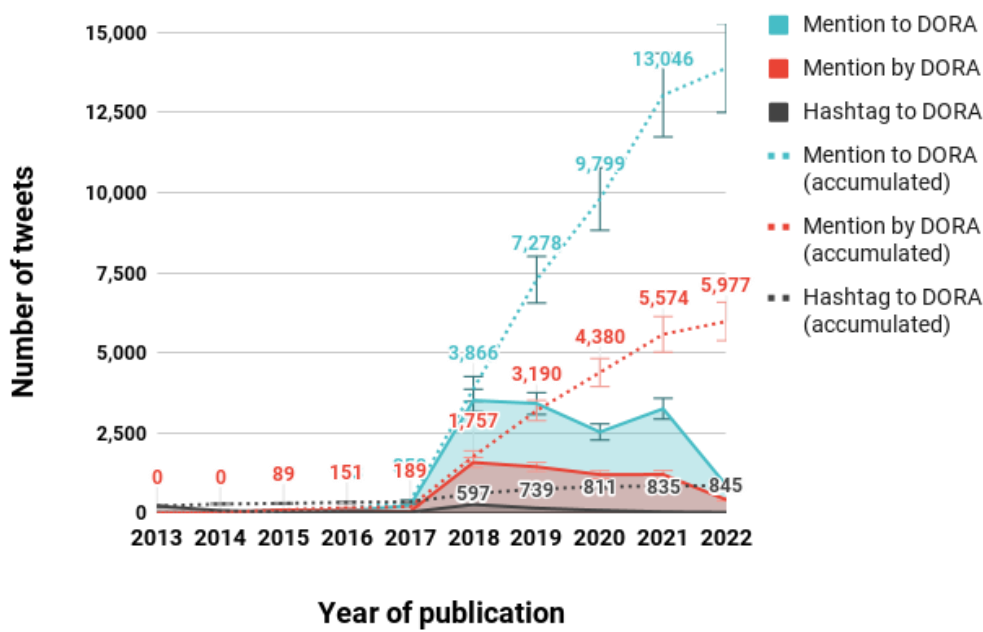

**Fig.1** Yearly and cumulative number of tweets about DORA (April 2015 to May 2022)

Most tweets about DORA are written in English (83%), with a remarkable presence in Spanish (6.5%). Specifically, DORA's account is published mainly in English (91.3% of all tweets), as this is the language used officially by DORA to publicize and disseminate their activities. The user-mention dataset shows a lower predominance of English (79.3%) and a higher presence of other languages, especially Spanish (8.1%) and Dutch (2.3%), covering users writing in their vernacular languages. Finally, most of the tweets in the hashtag-mention dataset are in English (85.0%), with a significant percentage of tweets in French (5.7%), most of them related to the adherence of the *Institut National de la Recherche Agronomique* (INRA) to DORA. Appendix C includes additional data on the presence of languages in the tweets.

**RQ2. Who is discussing DORA on Twitter?**

The hashtag-mention dataset comprises 419 users, of which 77.3% have published only one tweet, constituting a small set of users (see Appendix D for additional data). The user-mention dataset comprises 3,432 users, of which 54.39% have published only one tweet mentioning DORA's Twitter account. In comparison, a few (10 users) have posted more than 100 tweets, revealing their active role on the topic.

The users who mention DORA most frequently are compiled in Table 1, along with some descriptive data (total tweets, number of followers, account type, user role, and genre). Among these most productive users, we highlight the presence of Stephen Curry (chair of DORA) and Elizabeth Gadd (Vice-Chair of the Coalition on Advancing Research Assessment- CoARA), who use Twitter as loudspeakers to promote DORA. The presence of institutional accounts is scarce. Redalyc (*Red de*

*Revistas Científicas de América Latina y El Caribe, España y Portugal*) is the most productive institutional account, followed by the Open Research Leeds (64 tweets) from Leeds University Library and the scholarly publisher Queios (36 tweets). The presence of practitioners (e.g., Jeroen Bosman or Elizabeth Gadd) is also noteworthy, reflecting the importance of DORA in the professional area of science management.

**Table 1** Twitter users who mention DORA's Twitter account most frequently

| User | Tweets | Type | Role | Genre | All tweets | Followers |
|---|---|---|---|---|---|---|
| Christopher Jackson | 378 | Personal | Researcher | Male | 111,900 | 36,681 |
| Stephen Curry | 363 | Personal | Researcher | Male | 79.100 | 21,792 |
| Jeroen Bosman | 284 | Personal | Practitioner | Male | 25,300 | 6307 |
| Elizabeth Gadd | 217 | Personal | Practitioner | Female | 18.500 | 8,664 |
| Björn Brembs | 196 | Personal | Researcher | Male | 52,300 | 10,701 |
| Anna Hatch | 183 | Personal | Practitioner | Female | 6,981 | 1,368 |
| Bianca Kramer | 141 | Personal | Practitioner | Female | 19,100 | 5,818 |
| Isidro Aguillo | 141 | Personal | Researcher | Male | 61,300 | 10,528 |
| *Redalyc* | 139 | Institutional | Database | N/A | 25,500 | 33,922 |
| Egon Willighagen | 136 | Personal | Researcher | Male | 113,400 | 5,223 |

Note: total tweets and followers counts as of 6 June 2023.

48.9% (1,791) of all unique users mentioning the DORA account or using DORA-related hashtags were identified as individual scholars in the open dataset of scholars on Twitter, reflecting how the conversation about DORA has attracted academics. The remaining users are individuals not identified as scholars, institutional accounts, and the public.

Otherwise, 6,103 unique users have been mentioned in DORA-related tweets, of which 5,613 have been co-mentioned with the DORA's official account but not directly mentioned by DORA. Among the most mentioned users, five of the most productive users (Stephen Curry, Christopher Jackson, Elizabeth Gadd, Anna Hatch, and Jeroem Bosman) stand out as the most influential Twitter actors about DORA (Table 2). In this case, we find a more significant presence of institutional accounts, including a university (University College London), a research council (NOW NIeuws), a research society (The Geological Society), an institutional repository (Arxiv), and a charitable foundation (The Wellcome Trust) in the Top 15 most mentioned users.

**Table 2** Twitter users most frequently mentioned (DORA and user-mention datasets)

| User | Type | Total mentions | Community Mentions | DORA mentions |
|---|---|---|---|---|
| Stephen Curry | Individual | 1,731 | 1,529 | 202 |
| Christopher Jackson | Individual | 915 | 859 | 56 |
| Elizabeth Gadd | Individual | 816 | 785 | 31 |
| Anna Hatch | Individual | 673 | 559 | 114 |
| Jeroen Bosman | Individual | 543 | 527 | 16 |
| Tanvir Hussain | Individual | 524 | 520 | 4 |
| *University College London* | Institutional | 502 | 470 | 32 |
| David Price | Individual | 437 | 434 | 3 |
| *NWO Nieuws* | Institutional | 429 | 394 | 35 |
| The Geological Society | Institutional | 408 | 406 | 2 |
| *Arxiv* | Institutional | 402 | 399 | 3 |
| Jennifer L. Rohn | Individual | 397 | 394 | 3 |
| Anson Mackay | Individual | 380 | 378 | 2 |
| Ying Lia Li | Individual | 374 | 373 | 1 |
| *Wellcome Trust* | Institutional | 372 | 299 | 73 |

A network of Twitter users mentioning and being mentioned is included in the supplementary material (Appendix E).

**RQ3. Which topics about DORA generate interest on Twitter?**

*Tweets categories*

The messages referring to DORA as an organization and DORA as the Declaration must be distinguished. DORA as an organization includes tweets about organized events where DORA has participated or been discussed (Tweet-ID: 1374727844765519882), prizes awarded (1496868703601758212), congratulations and celebrations (992071747137622017), collaborations (1344738751629713411), and material created, such as toolkit of resources (1372171862453665793) or case studies (1365317404696453126).

Among the DORA tweets as Declaration, the following categories have been distinguished:

*Adherence to DORA*

This category includes Tweets announcing that DORA has been signed, including individuals (1214937813126467586), journals (1143586732933300226), universities (1504129642835460103), centers (1036584670106927107) or companies (1096492803486564352). In the same way, we find tweets in which users report/consult that other institutions have (or have not) signed DORA (1372667819544997896) and also express complaints (1503688592899678218), surprises (1374784968635846663), signature requests (1408728924205506560), wishes (668321880701403136) or recommendations (1410363218271277060). Informative tweets indicating the number of signatory institutions or countries are also included in this category (1407328955405897730).

Critical tweets appear warning of hypocritical signatures. This way, journals that have signed DORA but advertise products contrary to the DORA principles (1229501070105337860) or show the Journal Impact Factor (JIF) as a claim (791264494643142661) are criticized. Likewise, universities that, having adhered to DORA, continue to use the JIF in promotions (1229487498751025152) or public communications (1329154315408248836) are complained, as shows DORA as a promotional gesture (1193420947036000256) or simple declaration of intent (1193420947036000256). This leads to the question of DORA's official reaction in these situations (1367207353653727233) and whether resignation is advisable (1182664394641215488).

*Informative tweets*

This category includes tweets that explain or clarify the objectives of DORA (1410446201359671298) and its intention to avoid the misuse of metrics (1399772027477704710). Additionally, tweets comparing DORA with other manifestos, such as Hong Kong Principles (1230836366046527488) or the Leiden Manifesto (971128376344924160), are also included.

*DORA support*

This category includes tweets that support or recommend the use of DORA for different reasons, among which we find the need to change incentives (1370794141416894467), avoid the use of rankings (1381111833050800128), and support Open Science (1373895972317663233).

*DORA alignment*

This category includes tweets that discuss the greater or lesser alignment to DORA of specific research and academic activities, including publishers' policies (1368599813185544200), databases (1381335472090177536), or recruitment procedures (1500903257748738049). This category also includes tweets in which activities related to Open Science are valued, such as support for the preprint (1400525600893259777) or the narrative resumés (1346388818124005377).

*DORA implementation*

This category includes tweets related to the actual implementation of DORA. Some users comment on the difficulty of this implementation when trying to modify the firmly established behavior of the community (1105467868559360000), advising that it still requires effort (1233409691570573315), strong strategy (1323608595360030720) and that the reward system must change first (1105424140239020032). Other users are more critical, indicating that DORA is just well-wishing (126360194914600141) and is not working (483518611731283969). There is also some fear that DORA will be reduced to believing that banning the JIF would mean achieving responsible metrics (995247824295092229).

Finally, other users warn of the need for DORA to be proactive towards institutions that violate DORA's principles after having joined to expel them (965512696362950656), which could give DORA greater credibility in the community, encouraging their implementation.

*DORA limitations*

This category includes those tweets that highlight perceived limitations of the DORA principles.

For example, we find tweets in which alleged contradictions are pointed out, like a user who stresses concern about the journal in which an article is published. Still, elitist criteria are established regarding the press where books are published (1199623958389231616). In other cases, the fact that article-level metrics are recommended when these are not accurate is criticized. Users also complain that banning the JIF of their publications in their CVs is a non-transparent top-down exercise since authors should be able to show all the metrics they deem appropriate (1455864229295280129).

*Qualitative and quantitative methods*

This category is related to the DORA implementation and integrates critical tweets with the qualitative approach on which DORA is based.

For example, some users reject being evaluated with qualitative criteria (1484117976945156102) and even claim which specific indicators would be used (1484076951891591168). Doubts are also concerned about the potential subjectivity of the qualitative evaluation since i) it would give all the power to the committees, which do not necessarily have the expertise to make these decisions (1450459093827735552); ii) it does not provide a realistic solution when generating a high workload for evaluators (1174246251442900992); and iii) it makes difficult to compare CVs (1212333495944654848). Likewise, the possibility of not accessing expert reviews could pose another problem (1450488406316920838).

*Use of bibliometric indicators*

This category includes those tweets focused on using bibliometric indicators and the debate around the JIF.

Taking apart those tweets against using the JIF (959465383412584449), other users recognize its usefulness in identifying quality journals when the journal is not previously known (1073219403360006144). Likewise, it is indicated that it is easy to criticize the JIF, but a solution is not offered (1009076256707620865), and that badges do not better represent the merit of an article (1096107025660932097). In other cases, the refusal to use the JIF as the only individual measure is assumed, but the total exclusion of journal metrics is extreme (1435954102928920578).

In line with the appropriate use of the indicators, there is also a debate about the need for Bibliometrics experts in the evaluation commissions, with conflicting positions in this regard (1483905847063416834).

*DORA effects*

This category includes those tweets focused on discussing the possible effects of implementing DORA in the scientific community.

Regarding the changes in the evaluation of science in general and of DORA in particular, a few users indicate that many speak, but few do (1331146598898937858). Although they recognize that DORA is a positive initiative, they also express that it can generate more problems than solutions (1212329318992494593) and even polarize researchers (1105447238162178048). In other cases, there are doubts about their actual impact on researchers (1032943306488139778), their ability to eliminate the influence of journals (1376911302732439558) or mitigate other actions (1352686004377964545). Other users indicate that DORA and similar actions are initiatives that cannot last (1367196622665433090) or are simply gestures (382974286308204544).

Other users indicate that research evaluation would be chaotic without a clear idea of how research performance will be measured (1009512744888995840). Also, the lack of clarity regarding what is expected from faculty members would cause stress (1236572887756279808), or assessing research on its merit leaves much room for ambiguity (1149342343939538944).

*DORA awareness*

This category includes those tweets focused on discussing the dissemination of DORA.

We find tweets where users narrate when or how they discovered DORA (1141247756830498816), regret how many researchers are still unaware of this movement (857223222319869952), or complain about news reporting actions contrary to DORA (836990428192849920).

*Hashtags categories*

Twenty categories have been detected, plus an additional category (undetermined) for those hashtags with no precise topic (Table 3). These categories are aligned with the categories of tweets previously described. We can observe a strong presence of Assessment (1,518 tweets), Declarations (1,260), and Open (1,096). We also detect the use of specific topics related to research assessment, such as the specificity of fields of knowledge (184), the consideration of gender and diversity (122), the need for ethics (77), or the dependence on data sources (77).

**Table 3** Hashtag categories

| Category | Number of tweets | Sentiment (avg.) | Sentiment (max.) | Sentiment (min) | Scope |
|---|---|---|---|---|---|
| Academic life | 904 | 1.43 | 10 | -6 | Hashtag related to the research activity expressed in general terms (e.g., #science). |
| Assessment | 1,518 | 1.62 | 10 | -5 | Hashtag related to research evaluation activities (e.g., #evaluation). |
| Community | 187 | 1.45 | 8 | -10 | Hashtag related to individuals or people (e.g., #researcher, #scholars). |
| Data sources | 77 | 1.51 | 10 | -10 | Hashtag related to databases (e.g., #scopus, #webofscience). |
| Declarations | 1,260 | 1.34 | 10 | -7 | Hashtag related to a Statement, Declaration, or Principle about research assessment (e.g., #DORA, #leidenmanifest). |
| DORA Support | 122 | 0.81 | 6 | -2 | Hashtag that explicitly supports DORA (e.g., #RedalycJournalsSignDORA). |
| Ethics | 77 | 2.45 | 10 | -5 | Hashtags related to the science and research practices (e.g., #fakescience, #researchethics). |
| Events | 867 | 1.29 | 10 | -6 | Hashtag related to a scientific or professional meeting (e.g., #RRAConference, #SNFSConference). |
| Expression | 36 | 1.00 | 7 | -6 | Hashtag related to a linguistic term or jargon (e.g., #DYK, #TBT). |
| Field/Discipline/Subject | 184 | 1.68 | 10 | -4 | Hashtag related to the different branches of science (e.g., #humanities, #maths). |
| Funding | 73 | 1.29 | 6 | -2 | Hashtag related to research funding activities or funding projects (e.g., #ResearchFunding, #horizoneurope). |
| Gender, diversity, equality & inclusion | 85 | 1.55 | 6 | -2 | Hashtag including terms related to inclusive topics (e.g., #inalllanguages, #inclusiveassessment). |
| Institutions | 262 | 1.57 | 10 | -3 | Hashtag that mentions entities in general (e.g., #libraries, #department) or specific (e.g., #coalitionS, #CardiffUni) terms. |
| Location | 70 | 1.64 | 10 | -5 | Hashtag mentioning a place, including cities, countries, continents, or regions (e.g., #eeuu, #helsinki). |
| Metrics | 903 | 1.92 | 10 | -6 | Hashtag mentions metrics (e.g., #noimpactfactor). |
| Open | 1,096 | 1.32 | 10 | -8 | Hashtags related to the open movement: open access (e.g., #DiamondOA), open data (e.g., #datasharing), or open science (e.g., #opensci). |
| Peer review | 62 | 1.33 | 10 | -5 | Hashtag related to the peer-review process (e.g., #postpublicationreview). |
| Position/Career/Award | 78 | 1.47 | 9 | -2 | Hashtag related to research position (e.g., #postdoc), academic career (e.g., #ResearchCareers), or recognition (e.g., #MaddoxPrize). |
| Publication | 356 | 1.26 | 10 | -5 | Hashtags related to the academic publication system, including journals and publishers (e.g., #PLOS, #Preprint). |
| Software & Technology | 41 | 2.09 | 8 | -1 | Hashtag mentioning applications (e.g., #wordpress) or technologies (e.g., #RoR, #blockchain). |
| Undetermined | 214 | 1.42 | 10 | -3 | Hashtags that could not be categorized nor have a specific meaning (e.g., #27August). |

*Co-occurrence of terms*

The clusters identified from the co-occurrence of terms included in the tweets are shown in Fig.2. These clusters summarize the following ten topics of discussion (order by the number of nodes): "Open science and research assessment" (32 nodes); "Academics career assessment & innovation" (14 nodes); "Journal Impact Factor" (11 nodes); "Outputs scientific research" (9 nodes); "Individual assessment & quality" (7 nodes); "Impact metrics" (7 nodes); "Research outputs & influence" (7 nodes); "San Francisco Declaration" (5 nodes); "Scientific content and committees making decisions" (5 nodes); and "INRA signé Declaration" (3 nodes). Appendix F includes information for all identified clusters (e.g., label nodes, cluster name, weight). Complementarily, a co-occurrence network of hashtags has been performed (Appendix G).

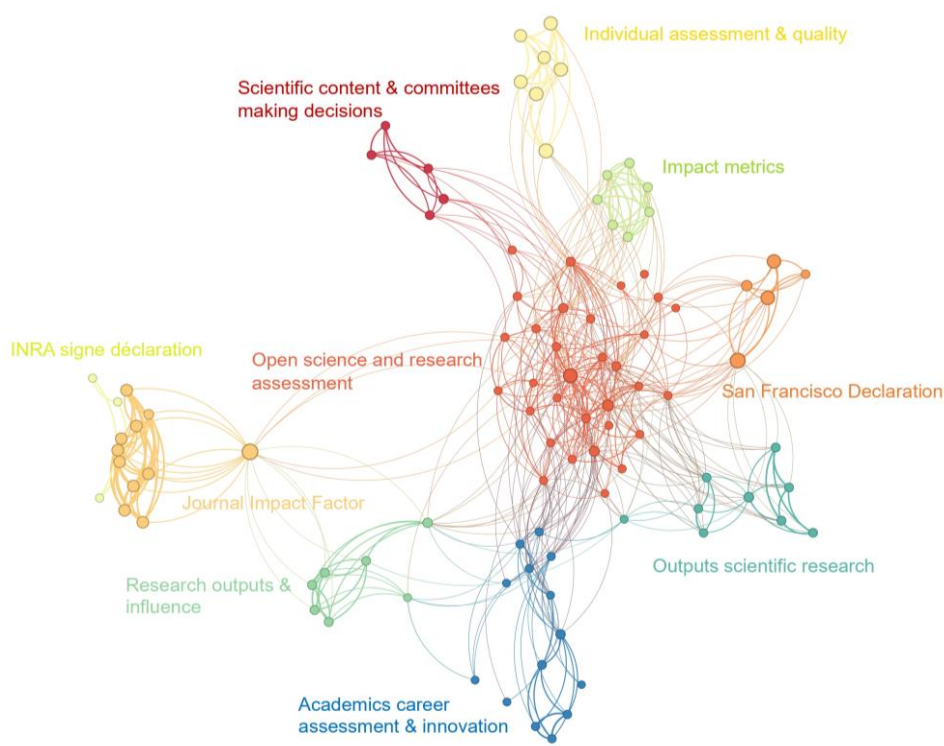

**Fig.2** Topics of discussion about DORA through a co-occurrence network of terms
Note: terms extracted with CorTexT (100 most frequent keywords and specificity computed using the chi2 score).
Map created with Gephi 0.10.1 (OpenOrd distribution). Available at:
https://documents.cortext.net/lib/mapexplorer/explorerjs.html?file=https://assets.cortext.net/docs/1389750259a166f5af00b76777be843d

**RQ4. Do topics about DORA on Twitter evolve?**

The evolution of topics over time (2015-2022) is shown as a tube layout graph (Fig.3). Individual tube layouts are available in Appendix H. Results show an upward trend in the discussions (from two topics in 2015 to seven topics in 2021). During the initial period (2015-2017), discussions were focused on research metrics, impact factor, and the Declaration in general.

We can highlight three critical tubes. The first (in red) is related to research assessment and research evaluation, with open science becoming predominant since 2019. The second (also in red) refers to measuring impact and indicators (with a decreasing trend since 2019). The third topic (in green) comprises research and career assessment reforms, which have gained prominence since 2020.

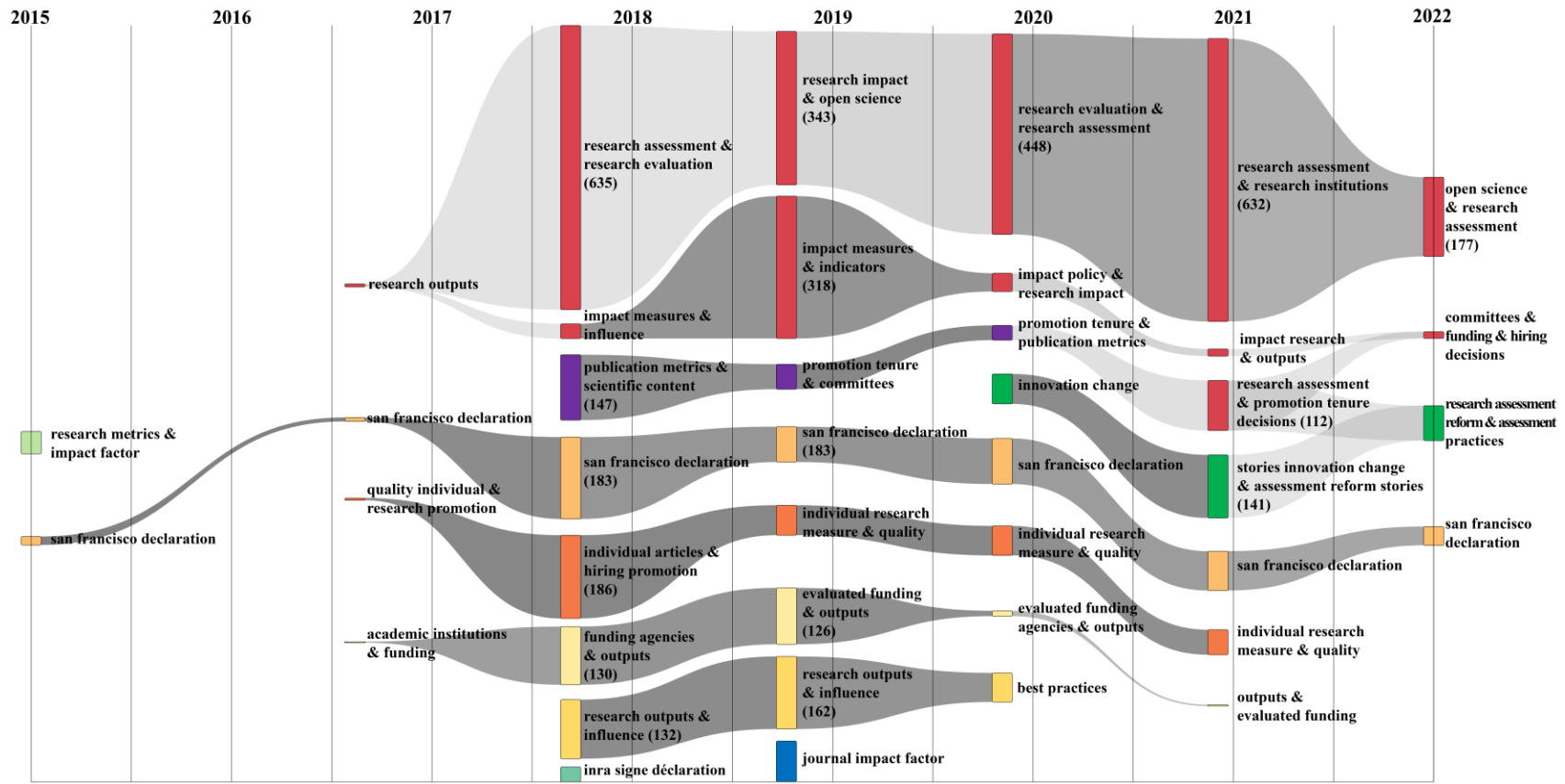

**Fig.3** Sankey diagram of the evolution of topics over time
Note: The timeline covers information until 31 May 2022. All three datasets are included. Numbers in parentheses indicate the number of tweets.

**RQ5. What sentiment do the Tweets about DORA generate?**

Regarding the datasets, tweets published by DORA show a positive monthly average sentiment (1.32) (Fig.4, top), which seems plausible as this account would provide informative (and positive) information about the Declaration. The lower monthly average corresponds to June 2020, with a neutral value of 0, while the maximum (1.9) corresponds to February 2020. The tweets from the user-mention dataset show a lower but positive average (0.92), with a broader statistical range during the first years (Fig.4, middle). The highest negative sentiment score (-6) was found in March 2017. Tweets from the hashtag-mention dataset show a higher positive average score (2), having a maximum score of 8 (April 2020) and a minimum of -2 (November 2016). The monthly average sentiment analysis raw scores for each dataset are available in Appendix I.

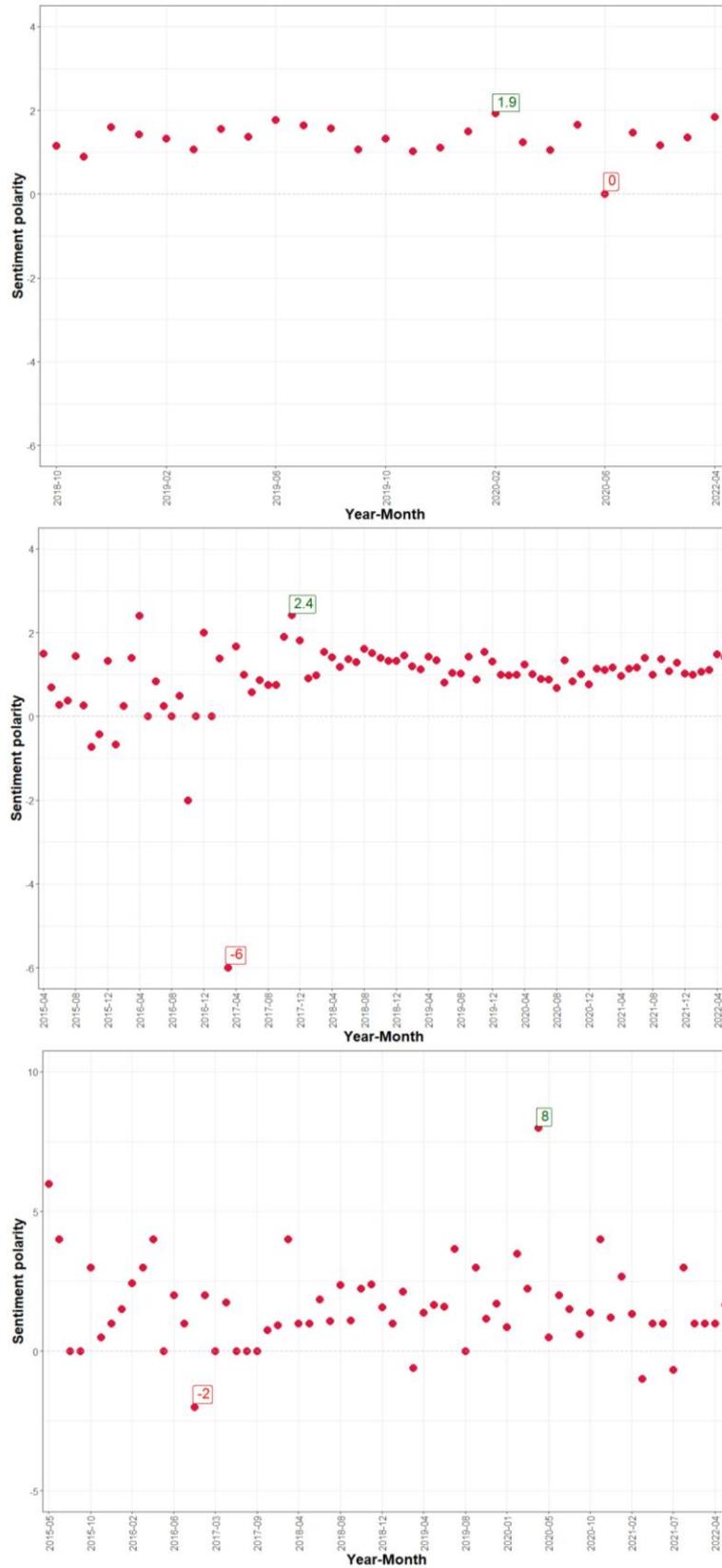

**Fig.4** Monthly average sentiment score for the DORA (top), user-mention (middle), and hashtag-mention (bottom) datasets

The sentiment of tweets vary according to their category. In this sense we can observe how those tweets related to "Ethics" and "Software and technology" obtain, on average, higher sentiment scores (Table 3). However, the number of tweets in these categories is low (77 and 41, respectively). In contrast, tweets related to "DORA Support" receive the least positive sentiment on average. The average sentiment on the remaining categories oscillates between 1.3 and 1.6. The higher negative sentiments are found with "Community" and "Data Sources" (the maximum value of -10 is achieved), followed by "Open" (-9), which appear as the most controversial topics.

**RQ6. What output-level impact do the tweets about DORA generate?**

Around 69% of the tweets published by DORA have received at least one like observing a similar pattern for the user-mention dataset and a lower impact in the hashtag-mention dataset (55% of these tweets did not receive likes). We can observe differences between the three datasets considering the remaining impact metrics. Tweets published by DORA are highly retweeted but scarcely replied to; this pattern is like that achieved by tweets including DORA-related hashtags. However, the tweets mentioning DORA are scarcely retweeted but highly replied to. Finally, the percentage of quoted tweets is scarce in all datasets, especially for the user-mention dataset (Table 4).

**Table 4** Number of tweets with no impact

| Metrics | DORA dataset | | User mention dataset | | Hashtag mention dataset | |
| --- | --- | --- | --- | --- | --- | --- |
| | No. Tweets | % | No. Tweets | % | No. Tweets | % |
| Likes | 1,902 | 31.82 | 4,339 | 31.23 | 463 | 54.8 |
| Retweets | 569 | 9.52 | 9,747 | 70.15 | 317 | 37.5 |
| Replies | 5,471 | 91.53 | 7,813 | 56.23 | 797 | 94.3 |
| Quotes | 4,552 | 76.16 | 12,633 | 90.92 | 691 | 81.8 |

Regarding output-level impact data intensity (Fig.5), quotes and replies counts achieve low values. Otherwise, neither likes nor retweets reach outstanding values (e.g., the tweet receiving the most likes attracted 523 likes, whereas the tweet receiving the most retweets attracted 483 retweets). If we focus on the number of likes, the results show 29 tweets with at least 100 likes (20 in the user-mention dataset, 16 in the DORA dataset, and 3 in the hashtag-mention dataset), which evidence a low output-level impact.

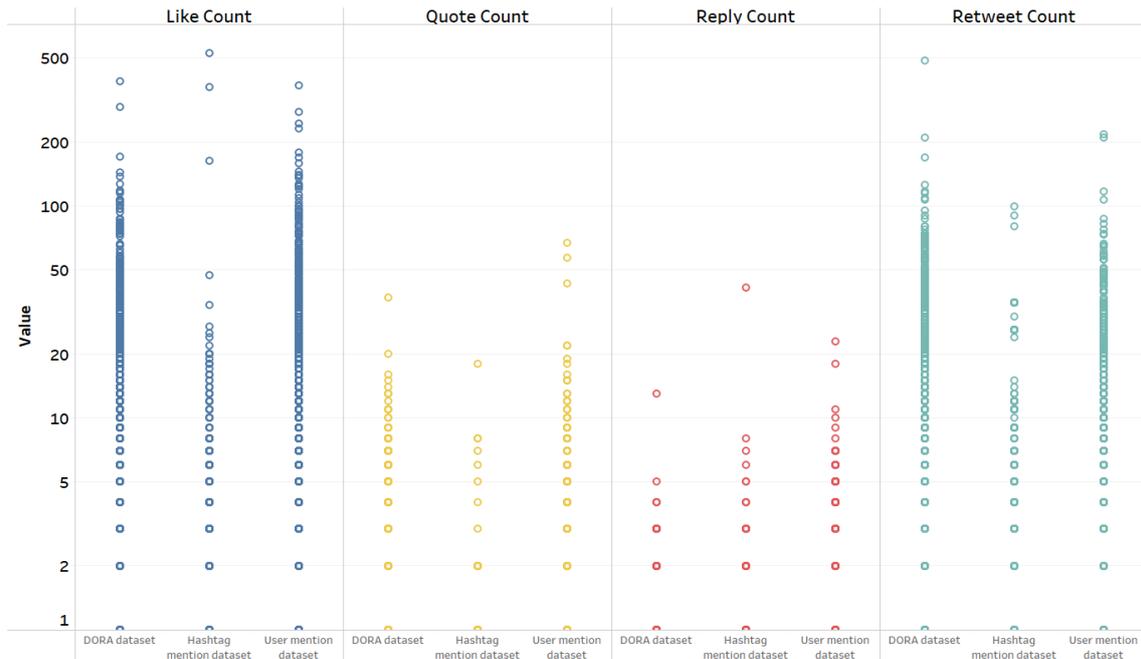

**Fig.5** Distribution of impact metrics (likes, retweets, replies, quotes) per dataset

The output-level impact (number of likes) and the average sentiment are integrated to monitor the interest of each hashtag category over time (Fig.6). This way we can observe specific bursts in Publication (Q4, 2017), Peer review (Q4, 2018; Q1, 2022), Open (Q3, 2021), Location (Q3, 2021), Gender (Q4, 2021), Ethics (Q3, 2021), Declarations (Q3, 2021), and Community (Q3, 2021) categories. Overall, the average sentiment is skewed toward positive comments (especially Ethics), while negative comments are seldom. In an case, as each dot represents a three-month window, extreme values of specific tweets might be masked. The supplementary material (Appendixes J and K) includes detailed data on all the social media metrics analyzed.

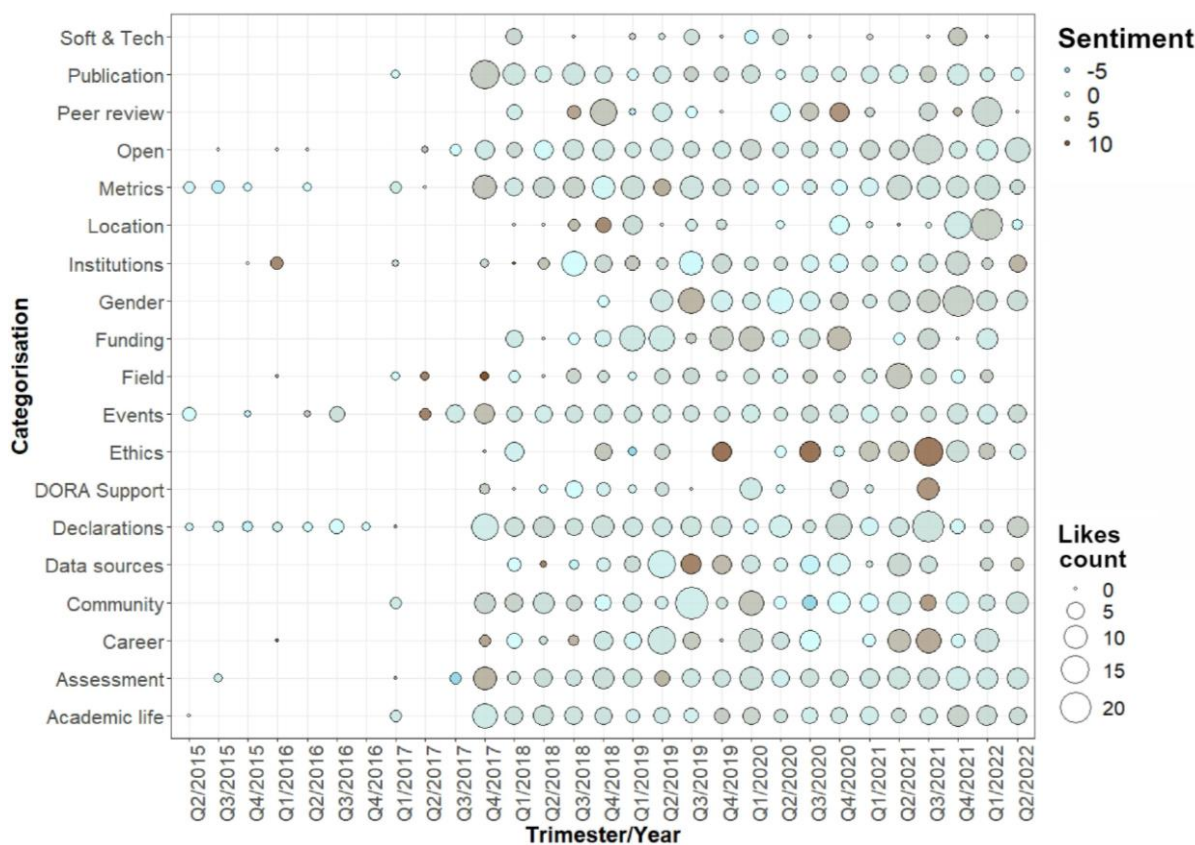

**Fig.6** Trimester evolution of sentiment analysis by hashtag category

## 5 Discussion

**Tweets volume and evolution**

A growth in the number of tweets published over time is observed, a trend also identified in the analysis of other scientific-related topics on Twitter like the h-index (Thelwall & Kousha, 2021) and Open Access (Sotudeh, 2023), which could indicate a growth in the scholarly use of Twitter to debate scientific topics.

Given the nature of the three datasets obtained, the authors consider that the volume of tweets collected (20,807) is significant for the total number of tweets about DORA, which is not estimated to exceed 30,000. Although the results are not comparable, Thelwall and Kousha (2021) collected 30,681 non-duplicate English-language tweets on the h-index, while Sotudeh (2023) collected 9,268 OA-related tweets published in 2019. These data lead us to believe that the volume of discussion about DORA is lower.

Although the number of tweets presents an upward trend, the discussion lagged until 2015, presumably related to the fact that the account was created three years later than the launch of the Declaration (2012). Other declarations (e.g., Hong Kong principles, Leiden Manifesto) still lack a Twitter account.

The maximum activity during the study period is 2018, with 5,333 tweets published (1,568 by DORA, 3,508 in the user-mention dataset, and 257 in the hashtag-mention dataset). This rise in tweets could be associated with two events launched that year: the announcement of Plan S[2] (Smits & Pells, 2022) and the creation of the European Open Science Cloud[3]. Similarly, the strong linkage between Open Science and DORA (Abadal, 2021) could explain this rise (e.g., hashtags assigned to the Open category rose from 1 in 2015 to 221 in 2019). The results evidence the DORA account acted as a seed of the debate on DORA on Twitter, leading to a multiplier effect by the Twitter community encompassing multiple talks.

**Twitter users**

Ten users have published more than 100 tweets mentioning DORA. Of these, individuals with an activist role (e.g., co-founder of Earth Arxiv, chair of DORA, or Vice-Chair of CoARA) stand out. This attests to the leading role of these key players, which concurs with previous studies of potential opinion leaders acting as influencers or science brokers (Díaz-Faes et al., 2019).

The set of users who have mentioned DORA or included tweets related to DORA has risen to 3,663 users. The authors estimate this value a low score, considering the general interest of the subject (research evaluation), the years of the period analyzed (8), and the number of actors involved (authors, journals, universities, research centers, etc.).

Furthermore, half of the users have been identified as researchers in the open dataset of scholars on Twitter. Considering the potential limitations of this dataset (mainly coverage and freshness), the authors estimate the presence of scholars as elevated. Conversely, the institutional participation in the debate around DORA is low beyond the promotional tweets supporting DORA.

**Tweets topics**

The corpus of tweets analyzed has reflected a broader debate around research evaluation, in which DORA participates. In fact, on many occasions, the mention of DORA occurred in response to a previously initiated discussion on research assessment (e.g., 1141286433870766081). This might explain why some of the most critical terms linked to open science, such as open peer review, open data (FAIR), preprints, or citizen science (Abadal, 2021), have not appeared among the most prominent discussion topics in the co-occurrence network (see Fig.2).

While the categorization of tweets made it possible to identify the top ten DORA-related categories, the co-occurrence of terms determined ten discussion topics where the interconnection of DORA with the evaluation of the research is most appreciated, including "Open science and research assessment," "Academic Career assessment & Innovation," and "Journal Impact Factor."

In a similar vein, the categorization of hashtags reinforced this tendency by highlighting Assessment (1,439 tweets), Declarations (1,115 tweets), Open (896), Metrics (835), and Academic Life (793) topics. All three methods were complementary to obtain an accurate picture of the discussions on Twitter.

The results have evidenced that the DORA-related tweets include those main concerns already expressed in scholarly publications (e.g., JIF refusal, qualitative review concerns, academic career assessment change) (Pérez-Esparrells, Bautista-Puig, Orduña-Malea, 2022). Twitter has been additionally able to locate DORA-related conversations such as events (e.g., institutions signing DORA, recognitions, awards) and critics (e.g., hypocrisy signing), offering greater plurality (horizontal vector) and granularity (vertical vector) in the study of the DORA phenomenon.

---

[2] https://www.coalition-s.org
[3] https://eosc-portal.eu

**Tweets sentiment**

The results evidence a moderately positive tone, exhibiting a more vital polarity in the first years (2015 to 2018), presumably related to the enthusiasm towards DORA in the scientific community. However, the prevalence of the neutral tone could indicate that tweets are mainly informative (e.g., an institution that is joining the Declaration or individuals explaining the principles of the Declaration), with a few negative exceptions (e.g., the concern of the individuals towards the ways they are evaluated). The brevity of the texts on Twitter could explain the neutrality of the tweets, as the longer the text, the higher the statistical probability of expressing sentiment.

The analysis of Tweets has identified messages for and against the DORA principles. However, the discussion has not generated content with an extreme sentiment but rather criticism, irony, and conflicting opinions, in which congratulatory messages for DORA signatures, negative messages towards using the JIF, and critics of hypocrisy DORA signatures prevail.

However, the results should be taken cautiously due to the difficulties of sentiment analysis when detecting the tone of messages. Moreover, this study has applied a dictionary method, which introduces technical limitations (Van Atteveldt, Van der Velden, & Boukes, 2021).

**Tweets output-level impact**

Although likes are the most prevalent engagement metric for measuring users' social media activity around science, around one-third of the dataset did not attain any. This result is aligned with previous studies that found engagement is more associated with factors such as the number of mentioned users in tweets or the number of followers (Fang et al., 2022). In this vein, tweets published by DORA achieve more likes; however, these tweets are not replied to or quoted, indicating a lack of conversation. A plausible explanation is that the content published is mainly informative, which aligns with the sentiment scores obtained. However, tweets posted by the Twitter community, especially on some topics such as open science, receive more engagement (i.e., are more retweeted).

**Limitations**

The analysis of Twitter includes a few limitations that should be acknowledged. The first limitation is data stability. As an example, users of social media could close their accounts or change their profiles, which might have affected the content retrieved from this source. The second limitation is data comprehensiveness. The search query might not include some results (e.g., the discussion on DORA does not mention the official account or uses any other name besides DORA, and the noise of the #DORA hashtag jeopardizes its use). The third limitation is the date range. Tweets were captured until 31 May 2022 and might not have captured the trend evolution (e.g., the DORA's 10[th] Anniversary and the associated events). Beyond Twitter, using CorTexT for identifying topics could not have reflected the whole spectrum. Finally, coding tweets and hashtags introduces an inherent subjectivity that should be considered.

# 6 Conclusions

Despite the results being confined to the Twitter universe, this platform has offered a granular and enriched vision of DORA, covering not only the main concerns around research assessment already covered in the literature but also other social aspects that help to understand the spread and interest of DORA in the research community.

Putting aside the merely promotional and informative tweets, it can be concluded that the debate around DORA on Twitter has brought together few participants (especially assessment agencies, funders, journals, or universities), has attained low interest, and has reflected a general moderate positive vision of DORA, probably because skeptical people have not fully participated, for different reasons.

The DORA debate has appeared as part of a broader debate (research evaluation), and the attention received is highly dependent on other related movements, especially Open Science. Analyzing other social platforms and similar declarations would confirm whether these results are due to the Twitter environment or reflect a more general feeling.

The results obtained may be of interest to research evaluation and Science studies scholars, to the promoters of DORA or other similar initiatives, as well as to the different evaluators and evaluated agents when it comes to knowing the interests and concerns of the scientific community around the reforms in the evaluation of science.

# 7 Supplementary Material

The supplementary material is openly available at the following link: https://doi.org/10.4995/Dataset/10251/199150

# 8 Funding